\documentclass[aps,pra,twocolumn,showpacs,superscriptaddress,bibliography]{revtex4-1}
\usepackage{graphicx}    
\usepackage{dcolumn}    
\usepackage{bm}             
\usepackage{amssymb}   
\usepackage{amsmath}    
\usepackage[caption=false]{subfig}
\usepackage{braket}
\usepackage{mathrsfs}
\usepackage{multirow}
\usepackage{dsfont}
\usepackage[export]{adjustbox}
\usepackage[toc,page]{appendix}
\usepackage{enumitem}

\begin{document}

\title{Trapped-ion two-qubit gates with $>99.99\%$ fidelity without ground-state cooling}

\author{A.~C.~Hughes}
\affiliation{Oxford Ionics (an IonQ company), Oxford, OX5 1PF, UK}
\author{R.~Srinivas}
\affiliation{Oxford Ionics (an IonQ company), Oxford, OX5 1PF, UK}
\affiliation{Department of Physics, University of Oxford, Oxford, OX1 3PU, UK}
\author{C.~M.~L{\"o}schnauer}
\affiliation{Oxford Ionics (an IonQ company), Oxford, OX5 1PF, UK}
\author{H.~M.~Knaack}
\affiliation{Oxford Ionics (an IonQ company), Oxford, OX5 1PF, UK}
\author{R.~Matt}
\affiliation{Oxford Ionics (an IonQ company), Oxford, OX5 1PF, UK}
\author{C. J. Ballance}
\affiliation{Oxford Ionics (an IonQ company), Oxford, OX5 1PF, UK}
\affiliation{Department of Physics, University of Oxford, Oxford, OX1 3PU, UK}
\author{M.~Malinowski}
\email{mm@oxionics.com}
\affiliation{Oxford Ionics (an IonQ company), Oxford, OX5 1PF, UK}
\author{T.~P.~Harty}
\affiliation{Oxford Ionics (an IonQ company), Oxford, OX5 1PF, UK}
\author{R.~T.~Sutherland}
\affiliation{Oxford Ionics (an IonQ company), Oxford, OX5 1PF, UK}

\date{\today}

\begin{abstract}
We introduce the ``smooth gate'': a novel entangling gate method for trapped-ion qubits where residual spin-motion entanglement errors are adiabatically eliminated by ramping the gate detuning. We demonstrate the power of this technique by performing electronically controlled two-qubit gates with an estimated error of $8.4(7) \times 10^{-5}$ without the use of ground-state cooling. We further show that the error remains $\lesssim 5 \times 10^{-4}$ for ions with average phonon occupation numbers of up to $\bar{n} = 9.4(3)$ on the gate mode. These results show that trapped-ion quantum computation can be performed with high fidelity at temperatures above the Doppler limit, allowing for significantly faster and simpler device operation.

\end{abstract}
\maketitle

\section{Introduction}

The two-qubit entangling gate is a key building block of universal quantum computers (QCs) \cite{divincenzo_1995,nielsen_2010}, and typically the most difficult to realize at error rates commensurate with quantum error correction thresholds \cite{knill_1998,terhal_2005,aliferis_2005,hirota_2012}. Trapped ions are widely regarded as one of the most promising QC platforms, in part due to their continued ability to improve on two-qubit gates over the last two decades \cite{leibfried_2003,benhelm_2008,ballance_2016,gaebler_2016,srinivas_2021,clark_2021,loschnauer_2024}. However, existing methods rely on careful calibration of multiple control fields, and ground-state cooling of the ions' motion, to achieve high fidelities. For large-scale QCs, it is valuable to develop high-fidelity gate methods with relaxed control and temperature requirements.

Needing to ground-state cool in order to reach high two-qubit gate fidelities has placed a significant burden on quantum charge-coupled device (QCCD) trapped-ion architectures \cite{wineland_1998,kielpinski_2002,pino_2021}, with this cooling requiring over an order-of-magnitude more circuit time than gates \cite{moses_2023}. While two-qubit geometric phase gates \cite{molmer_1999,molmer_2000,leibfried_2003} can, in principle, achieve high fidelities at non-zero mode occupation, they still contain error mechanisms that worsen with temperature \cite{sutherland_2022_1}. As a result, all previous published demonstrations of high-fidelity gates in trapped ions involved ground-state cooling; not doing so would have resulted in non-trivially higher error rates. 

With laser-based gates, the prospect of eliminating this temperature sensitivity appears challenging. While some early work suggested adiabatic methods could greatly suppress residual spin-motion entanglement \cite{cirac_2000,calarco_2001,vsavsura_2003} -- one of the major temperature-dependent error mechanisms -- these techniques slow gates down, exacerbating the (typically dominant) error channels of photon scattering and phase noise. Additionally, higher-order Lamb-Dicke effects result in temperature-dependent errors that are difficult to avoid. 

Electronic two-qubit gates \cite{mintert_2001,ospelkaus_2008,ospelkaus_2011,harty_2014,harty_2016,weidt_2016,webb_2018,srinivas_2018,sutherland_2019,sutherland_2020,zarantonello_2019,srinivas_2021,barthel_2023,nunnerich_2025,loschnauer_2024} offer significantly more potential for temperature insensitivity. This is because (1) replacing lasers with near-field magnetic field gradients makes the Lamb-Dicke effects negligible, (2) low phase noise of radio- and microwave frequency sources allows flexibility in gate duration, and (3) the control fields do not couple to short-lived atomic states, making photon scattering negligible. Taking this into account, Ref.~\cite{sutherland_2024} suggested that using adiabatic methods similar to Refs.~\cite{cirac_2000,calarco_2001,vsavsura_2003} would render electronic gates extremely insensitive to motional frequency fluctuations -- possibly to the point they could maintain high fidelities without ground-state cooling. Prior to this work, however, this was only a theoretical prediction. \\

In this work, we demonstrate $\sim 0.9999$ (subspace benchmarked) two-qubit gate fidelities \textit{with no ground-state cooling}. This is, to our knowledge, the highest fidelity two-qubit gate ever demonstrated in any QC modality. This result improves by $\sim 3\times$ on the previous best reported fidelity of $0.9997$ \cite{loschnauer_2024} for ground-state cooled trapped-ion qubits, and $\sim 100\times$ on the best reported fidelity of $\sim 0.992$ without ground-state cooling \cite{srinivas_2020_thesis}. To do this, we develop a new gate scheme that adiabatically eliminates spin-motion entanglement by smoothly ramping the detuning of the gate drive with respect to the ions' motion, henceforth referred to as a ``smooth gate". The method avoids some of the implementation challenges and duration overheads that come from the amplitude ramping proposed in Ref.~\cite{sutherland_2024}, and provides significant motional robustness without the complexity of alternative coherent control methods \cite{hayes_2012, haddadfarshi_2016,shapira_2018,webb_2018}. 

The paper is structured as follows: In Sec.~\ref{sec:theory}, we provide a theoretical framework and use it to discuss the key differences between diabatic and adiabatic elimination. We also propose the smooth gate, and use our framework to derive some of its beneficial properties. Then, in Sec.~\ref{sec:experiment}, we present our experimental results, including (A) the experimental setup, (B) calibration procedures, (C) subspace benchmarking data, which we use to estimate gate fidelities at Doppler temperatures, (D) error rates versus static detuning offset, demonstrating the gate's motional robustness, and (E) infidelity versus starting temperature. Finally, in Sec.~\ref{sec:architecture} we discuss some of the architectural implications of our results.

\begin{figure*}[ht]
\includegraphics[width=\textwidth]{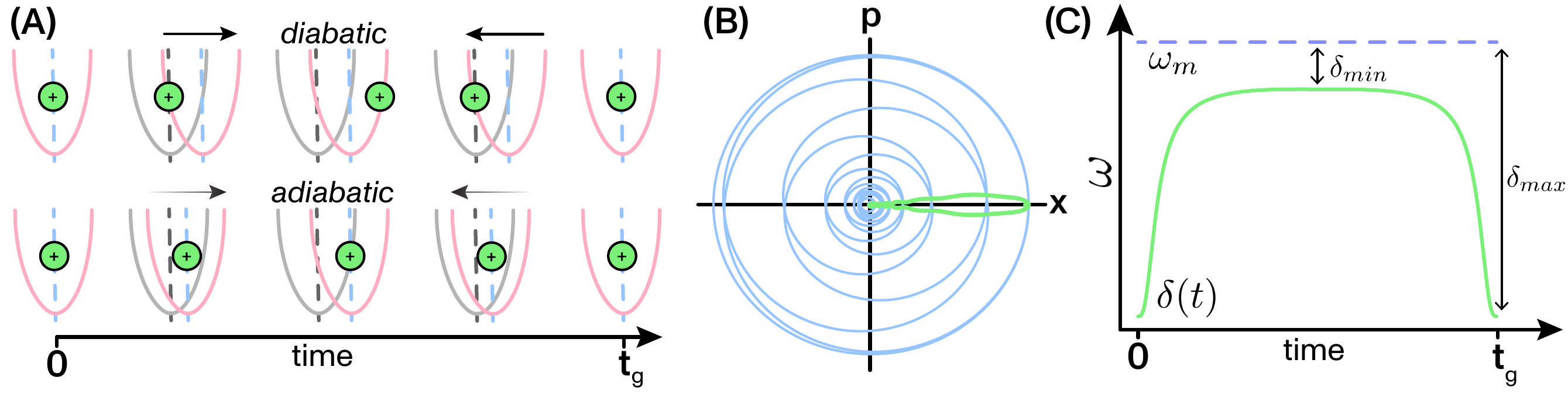}
\centering
\caption{A) Illustration of the dynamics for `diabatic' and `adiabatic' geometric phase gates. The pink (rightward) curves represent the well positions of the `forced' eigenstates and the grey curves represent the `null' eigenstates in the frame of Eq.~(\ref{eq:geo_general_lab}). Diabatic gates (top) rapidly turn on a spin-dependent force, causing the forced states to oscillate about new equilibrium positions. We `catch' the displaced motion at $t_{g}$ by rapidly turning off the spin-dependent force precisely when the motion returns to its initial state. For adiabatic gates (bottom), we slowly adjust the wells such that the displaced motion closely follows its instantaneous equilibrium position throughout the operation. B) Phase-space trajectory of an adiabatic gate in the frame of Eq.~(\ref{eq:geo_general_lab}) (green oval), and after transforming into the rotating frame with respect to the ions' bare harmonic motion (blue spiral) as described in Appendix~\ref{app:rotating_frame}. C) Frequency dynamics during a smooth gate. The dashed purple line is the gate mode frequency $\omega_{m}$ and the solid green line shows the frequency of the spin-dependent force.}
\label{fig:fig_1}
\end{figure*}

\section{Theory}\label{sec:theory}

 The most common method for implementing entangling gates in modern trapped-ion QCs is the geometric phase gate, which is based on spin-state-dependent forces acting on a set of harmonic oscillators. The spin-dependent force is in general detuned from resonance with the gate mode frequency $\omega_m$ by the gate detuning $\delta$. For two-qubit gates, we generate these forces using one or more frequencies $\omega_g$ of an electric or magnetic field gradient; for example, $\omega_g = \omega_0 \pm (\omega_m + \delta)$ for a M\o lmer-S\o rensen gate \cite{molmer_1999,molmer_2000}, where $\omega_0$ is the qubit frequency, or $\omega_g = \omega_m + \delta$ for a ZZ gate \cite{leibfried_2003, srinivas_2021}. The gate detuning may be time-dependent, and may be adjusted by changing either the gate mode frequency or the gradient frequency. Written in the interaction picture with respect to $\hbar(\omega_m-\delta)\hat{a}^{\dagger}\hat{a} + \hbar\omega_0\sigma_z/2$, the Hamiltonian takes the form:

\begin{eqnarray}\label{eq:geo_general_lab}
    \hat{H}_{g}&=& \hbar\delta(t)\hat{a}^{\dagger}\hat{a}+\frac{\hbar\Omega_g(t)}{2}\hat{S}_{\alpha}\Big(\hat{a}^{\dagger}+\hat{a}\Big),
\end{eqnarray}
where $\hat{a}(\hat{a}^{\dagger})$ are the motional mode's ladder operators, $\Omega_g(t)$ is the gate Rabi frequency, and $\hat{S}_{\alpha}\equiv \hat{\sigma}_{\alpha,1}\pm \hat{\sigma}_{\alpha,2}$ is a collective Pauli operator. Projecting Eq.~(\ref{eq:geo_general_lab}) onto the eigenbasis of $\hat{S}_{\alpha}$, we find two `forced' eigenstates (which experience a nonzero force), and two `null' states with eigenvalue 0. The forced states project the Hamiltonian onto:
\begin{eqnarray}\label{eq:forced_oscillator}
    \hat{H}_{f}&=& \hbar\delta(t)\hat{a}^{\dagger}\hat{a} \pm \hbar\Omega_{g}(t)\Big(\hat{a}^{\dagger}+\hat{a} \Big),
\end{eqnarray}
while the null states project it onto:
\begin{eqnarray}\label{eq:nonforced_oscillator}
    \hat{H}_{n}&=& \hbar\delta(t)\hat{a}^{\dagger}\hat{a}. 
\end{eqnarray}
Since each projected Hamiltonian is harmonic and independent ($[\hat{H}_{g},\hat{S}_{\alpha}]=0$), the system can thus be considered as four classical harmonic oscillators, two of them experiencing a force and two of them not. \\

The goal of a geometric phase gate is to implement an entangling interaction:
\begin{eqnarray}
    \hat{U}_{2q}&=&\exp\Big\{-\frac{i\theta_{g}}{2}\hat{S}_{\alpha}^{2} \Big\}.
\end{eqnarray}
For this to happen, two conditions must be met. First, the (time-dependent) force strength must be adjusted such that forced states accumulate the correct geometric phase, which is proportional to $\theta_g$. Second, the forced states and the null states must finish the gate at the same point in the $\braket{x}$ and $\braket{p}$ phase space -- a failure to accomplish this results in residual spin-motion entanglement. While residual geometric phase and residual spin-motion entanglement both lead to gate errors, the latter places more stringent requirements on ion cooling, as the magnitude of the associated gate error grows $\propto 2\bar{n}+1$, where $\bar{n}$ is the gate mode's average phonon number \cite{sutherland_2022_1}.

\subsection{Diabatic elimination of spin-motion entanglement}\label{sec:dese}

Historically, high-fidelity two-qubit geometric phase gates have operated at a constant gate detuning $\delta(t)=\delta_{g}$, and ramp $\Omega_{g}(t)$ on/off quickly relative to $1/\delta_{g}$ \cite{ballance_2016,gaebler_2016,srinivas_2021,clark_2021,loschnauer_2024}. For our `semiclassical' picture, this is equivalent to instantaneously shifting the equilibrium position of the forced states. Once $\Omega_{g}$ is on, the forced states oscillate about their new equilibrium positions, periodically returning to their original motional state (phase-space origin) every $t_{K}=2\pi/\delta_{g}$ (see top of Fig.~\ref{fig:fig_1}a). To eliminate residual spin-motion entanglement, we ramp off $\Omega_{g}$ at an integer multiple of $t_{K}$ -- `catching' the ions at the precise moment they return to their phase space origin. To contrast with gates that do this adiabatically, we refer to this as diabatic elimination of spin-motion entanglement (DESE). If $t_{g}$ is not an exact integer multiple of $t_{K}$, either due to a timing error or a drift in $\delta_{g}$, the system suffers spin-motion entanglement errors. Techniques such as Walsh modulation \cite{hayes_2012} are often employed to suppress these errors by reducing sensitivity to drifts in $\delta_g$ or miscalibrations of $t_g$.

At the same time, we must also calibrate the gate's entanglement angle $\theta_{g}$, which changes $\propto \Omega_{g}^{2}/\delta_{g}$. Since $\delta_{g}$ and $t_{g}$ cannot be tuned independently, the most straightforward way of calibrating $\theta_{g}$ is to scan $\Omega_g$. This, however, can be challenging in practice, as it involves changing the gate drive power, which can affect $\delta(t)$ (e.g. through changes in material temperatures), or cause qubit frequency errors through off-resonant AC Stark or AC Zeeman shifts. Thus, calibrating high-fidelity DESE gates involves a non-trivial amount of control complexity \cite{Gerster2022-qj, Yale2025}. In the subsequent sections, we show how smooth gates simultaneously minimise spin-motion entanglement errors and significantly ease the calibration challenge.

\subsection{Adiabatic elimination of spin-motion entanglement}\label{sec:adiabatic_cond}

In the QCCD architecture, quantum circuits are implemented by interleaving quantum gates with ion transport operations, such as split/merge, linear shifts, junctions, and swaps \cite{pino_2021,moses_2023}. Each of these can be thought of as a system of classical harmonic oscillators experiencing forces. In contemporary QCs, these operations are typically conducted in the `adiabatic' regime, meaning the trap potentials change slowly enough for each ion to closely follow its equilibrium position. This is for good reason: while diabatic transport can be faster, adiabatic transport is more robust, meaning cold movement can be achieved with relaxed control and calibration requirements. This is particularly important for large-scale systems, where it is beneficial to transport many ions using a single waveform \cite{burton_2023, malinowski_2023}.

Our method adapts this insight to two-qubit gates. Informally: since adiabatically turning on/off the state-independent forces responsible for ion transport is beneficial for robustly achieving low excitation, adiabatically turning on/off the state-\emph{dependent} forces responsible for geometric phase gates should be beneficial for robustly achieving low spin-motion entanglement. 

To achieve this `adiabatic elimination of spin-motion entanglement' (AESE), we do the same thing as adiabatic transport: adjust the equilibrium position $\Omega_g(t)/\delta(t)$ slowly relative to the oscillator frequency $\delta(t)$ (in the frame of Eq.~\ref{eq:geo_general_lab}). Approaching the AESE regime, the forced states follow their spin-dependent equilibrium positions with increasing devotion. So, when we ramp the spin-dependent force on/off at the beginning/end of the gate, the forced states begin/end with the same equilibrium positions as the null state\textemdash trivializing the task of eliminating residual spin-motion entanglement. As illustrated in Fig.~\ref{fig:fig_1}b (green), this leads the phase-space trajectory to become increasingly narrow in momentum space, converging to a line along the position axis in the AESE limit. Fig.~\ref{fig:fig_1}b also shows the phase-space trajectory in the rotating frame with respect to the motion (commonly used in the literature), where it spirals about the phase-space origin (see Appendix \ref{app:rotating_frame}). This phase-space behavior provides a general definition of an AESE gate, agnostic to the specific implementation.  

\subsubsection{Generalized derivation of AESE}\label{sec:adiabatic_cond}

We begin by transforming Eq.~(\ref{eq:forced_oscillator}) into a reference frame that follows the equilibrium position of the forced states. To do this, we use the displacement operator:
\begin{eqnarray}
    \hat{D}_{1}=\exp\Big(\pm i\alpha[t]\hat{p} \Big),
\end{eqnarray}
where $\alpha(t)$ is a real function corresponding to the equilibrium position at time $t$. Setting $\alpha(t)=\mp \Omega_{g}(t)/\delta(t)$ gives a Hamiltonian with no force terms:
\begin{eqnarray}\label{eq:adiabatic_first_frame}
    \tilde{H}_{f,1}&=& \hbar\delta(t)\hat{a}^{\dagger}\hat{a} - \frac{\hbar\Omega_{g}^{2}(t)}{\delta(t)} \mp \hbar\dot{\alpha}(t)\hat{p}.
\end{eqnarray}
The total forced-state time propagator is now:
\begin{eqnarray}\label{eq:first_order_adiabatic_u}
    \hat{U}_{f,t}=\hat{D}_{1}\tilde{U}_{f,1},
\end{eqnarray}
where $\tilde{U}_{f,1}$ is the time propagator of $\tilde{H}_{f,1}$. At the beginning/end of a gate, we ramp on/off the gradient fields, leading to $\Omega_{g}(0)=\Omega(t_{g})=0$, $\hat{D}_{1}(t_{g})\rightarrow \hat{I}$, and $\tilde{U}_{f,t}\rightarrow \hat{U}_{f,1}$. We now transform $\tilde{H}_{f,1}$ a second time, using:
\begin{eqnarray}
    \hat{D}_{2}=\exp\Big(\pm i\beta[t]\hat{x} \Big),
\end{eqnarray}
where $\beta(t)$ is again a real function, giving:
\begin{eqnarray}
    \tilde{H}_{f,2} &=& \hbar\delta(t)(\hat{a}^{\dagger}\mp i\beta[t])(\hat{a}\pm i\beta[t])-\frac{\hbar\Omega_{g}^{2}(t)}{\delta(t)} \nonumber \\
    &&-2\dot{\alpha}(t)\beta(t)\mp\hbar\dot{\alpha}(t)\hat{p}\pm i\hbar\dot{\beta}(t)\hat{p}.
\end{eqnarray}
When setting $\beta(t)=\dot{\alpha}(t)/\delta(t)$, this simplifies to:
\begin{eqnarray}\label{eq:adiabatic_equil_second}
    \tilde{H}_{f,2} &=&\hbar\delta(t)\hat{a}^{\dagger}\hat{a} - \frac{\hbar(\Omega_{g}^{2}[t]+\dot{\alpha}^{2}[t])}{\delta(t)}\pm i\hbar\dot{\beta}(t)\hat{p} \nonumber \\
    &\simeq & \hbar\delta(t)\hat{a}^{\dagger}\hat{a} - \frac{\hbar(\Omega_{g}^{2}[t]+\dot{\alpha}^{2}[t])}{\delta(t)}.
\end{eqnarray}
We neglected the diabatic term in the second line, assuming: 
\begin{eqnarray}\label{eq:adiabaticity_cond}
    \frac{\dot{\beta}(t)}{\delta(t)} \ll 1,
\end{eqnarray}
after which, $\tilde{H}_{f,2}$ does not produce residual spin-motion entanglement. Choosing smooth boundary conditions for the amplitude and detuning ramp functions, i.e. $\dot{\Omega}_{g}(0)=\dot{\Omega}_{g}(t_{g})=\dot{\delta}(0)=\dot{\delta}(t_{g})=0$, the second frame transformation vanishes as well ($\hat{D}_{2}\rightarrow \hat{I}$), making the final time propagator of the system:
\begin{eqnarray}
    \hat{U}_{f,t}&=&\hat{U}_{f,2}, \nonumber \\
\end{eqnarray}
and Eq.~(\ref{eq:adiabaticity_cond}) our final adiabaticity condition. Having satisfied this condition, Eq.~(\ref{eq:adiabatic_equil_second}) tells us that the forced states accumulate a phase (gate angle) of:
\begin{eqnarray}
    \theta_{g}&\simeq &\int^{t_{g}}_{0}dt^{\prime}\frac{\Omega_{g}^{2}(t^{\prime})+\dot{\alpha}^{2}(t^{\prime})}{\delta(t^{\prime})},
\end{eqnarray}
relative to the null states. Finally, as the diabatic correction $\propto \dot{\alpha}^{2}(t)/\delta(t)$ to $\theta_{g}$ is typically small, we will often write the approximate rate of entanglement generation in subsequent sections as simply: 
\begin{eqnarray}\label{eq:ent_rate}
    \dot{\theta}_{g} \simeq \frac{\Omega_{g}^2(t)}{\delta(t)}.
\end{eqnarray}

\subsection{Smooth gates}\label{sec:smooth}

AESE may be achieved in various ways. Ref.~\cite{sutherland_2024} proposed a scheme that maintains Eq.~(\ref{eq:adiabaticity_cond}) by keeping a fixed gate detuning $\delta(t)=\delta_{g}$ and ramping the gate Rabi frequency $\Omega_g(t)$ over a timescale $\tau_{g}\gg 1/\delta_{g}$. The scheme operates with AESE and will thus benefit from motional robustness, but using this method to reach errors of $\sim 10^{-4}$ faces several challenges. First, as discussed in Sec.~\ref{sec:dese}, tuning and ramping $\Omega_g(t)$ can lead to changes in mode and qubit frequency, creating additional error channels and calibration steps. Second, as the rate of entanglement generation is proportional to $\Omega_g^2(t)$ (Eq.~\ref{eq:ent_rate}), the AESE requirement to slowly ramp $\Omega_g(t)$ significantly slows down the gate.

To avoid these issues, the smooth gate we propose and demonstrate operates with AESE by instead ramping the gate detuning $\delta(t)$ while keeping $\Omega_{g}$ constant. Physically, this ramp can be achieved by ramping the gradient frequency or the motional mode frequency. This allows the adiabaticity condition in Eq.~(\ref{eq:adiabaticity_cond}) to be satisfied while maintaining a high entanglement generation rate (gate speed), and eliminates the need for precise tuning and ramping of $\Omega_g$.\\

\noindent The smooth gate has 5 basic steps:
\begin{enumerate}[label=\textbf{\arabic*})]
  \item With $\delta(t)$ at a maximum $\delta_{max}$, ramp $\Omega_{g}(t)$ to a maximum $0\rightarrow \Omega_{g}$ over a time $\tau_{g}\gg 1/\delta_{max}\ll t_{g}$.
  \item With $\Omega(t)=\Omega_{g}$ constant, ramp $\delta(t)$ to a minimum $\delta_{max}\rightarrow \delta_{min}$ over a time $\tau_{d}\sim t_{g}/2$.
  \item Keep $\Omega(t)=\Omega_{g}$ and $\delta(t)=\delta_{min}$ constant for a time $t_{c}$. This step is optional.
  \item With $\Omega(t)=\Omega_{g}$ constant, ramp $\delta(t)$ to a maximum $\delta_{min}\rightarrow \delta_{max}$ over $\tau_{d}$.
  \item With $\delta(t)$ at a maximum $\delta_{max}$, ramp $\Omega_{g}(t)\rightarrow 0$ over $\tau_{g}$.
\end{enumerate}
Note that we are free to combine step $1(4)$ with the beginning(end) of step $2(5)$ without changing the basic results.

\subsubsection{Choosing the ramp functions}

Our goal is to maximize the average value of $\dot{\theta}_{g}=\Omega_{g}^{2}(t)/\delta(t)$, implying we want to minimize $\tau_{g}$ and the average value of $1/\delta(t)$ without violating Eq.~(\ref{eq:adiabaticity_cond}). For the two amplitude ramping steps at the beginning and end of the gate, Eq.~(\ref{eq:adiabaticity_cond}) becomes:
\begin{eqnarray}\label{eq:amplitude_adiabaticity_cond}
    \frac{\ddot{\Omega}_{g}}{\delta_{max}^{3}}\ll 1,
\end{eqnarray}
suggesting we want a large $\delta_{max}$. During the two detuning ramp steps, the adiabaticity condition becomes:
\begin{eqnarray}\label{eq:detuning_adiabaticity_cond}
    \dot{\beta} &= &\frac{\Omega_{g}(\ddot{\delta}\delta - 3\dot{\delta}^{2})}{\delta^{5}} \ll 1,
\end{eqnarray}
suggesting we want to change $\delta$ quickly near $\delta_{max}$ and slowly near $\delta_{min}$. We choose our detuning ramp via a differential equation: 
\begin{eqnarray}
    \dot{\delta} &=& a\delta^{j+1}\sin^{2}(t/\tau_{d}),
\end{eqnarray}
which (approximately) accounts for this scaling and ensures smooth boundary conditions. This gives:
\begin{eqnarray}\label{eq:det_ramp_func}
    \delta(t) &=& (b+c g[t])^{-1/j},
\end{eqnarray}
where:
\begin{eqnarray}
    b&\equiv &\delta_{max}^{j} \nonumber \\
    c&\equiv &\frac{2}{\tau_{d}}\Big(\frac{1}{\delta^{j}_{min}}-\frac{1}{\delta^{j}_{max}} \Big) \nonumber \\
    g(t) &\equiv & \frac{t}{2}-\frac{\tau_{d}}{4\pi}\sin(2\pi t/\tau_{d}).
\end{eqnarray}
Numerically, we find that $j=3$ and $t_{c}=0$ tend to give the best trade-off between $t_{g}$ and adiabaticity. In Fig.~\ref{fig:fig_1}c, we plot $\delta(t)$ for  $j=3$, which shows $\delta(t)$ changing quickly near $\delta_{max}$, and significantly more slowly near $\delta_{min}$. We arrived at this $\delta(t)$ somewhat heuristically, comparing the powers of Eq.~(\ref{eq:det_ramp_func}) with a handful of other test functions that scale similarly; finding an optimal $\delta(t)$ is left to future work. \\

We will finish this section with several general observations about smooth gates. First, for a system with multiple motional modes, as long as $\delta(t)$ is chosen to maintain AESE with respect to the \textit{least} detuned mode, every other mode will be necessarily more adiabatic, thus will experience less residual spin-motion entanglement. Thus, smooth gates can be applied to multi-mode systems. Second, as long as $\delta_{min}$ is set to achieve AESE for $\theta_{g}=\pi/2$, we are free to decrease $\theta_{g}$ by increasing the value of $\delta_{min}$ -- again, this will only reduce the residual spin-motion entanglement. This property simplifies the gate calibration procedure (see Sec.~\ref{sec:calibration}), and allows for a straightforward implementation of continuously parametrized two-qubit gates. Third, the smooth gate technique can be applied to improve the robustness of all geometric phase gates, regardless if implemented using lasers (electric-field gradients) or electronics (magnetic-field gradients). That being said, the former approach suffers from significant thermally-induced fluctuations in $\Omega_g(t)$ \cite{wineland_1998, Cetina2022}, which are not suppressed by the smooth gate. Thus, the smooth gate is most beneficial for laser-free gates.

\subsubsection{Motional robustness of the smooth gate}\label{sec:smooth_robust}

To examine the robustness of the smooth gate to detuning fluctuations that lead to spin-motion entanglement errors, we start by simulating the gate in the presence of a small, oscillatory mode frequency shift of magnitude $\varepsilon$:
\begin{eqnarray}
    \hat{H}_{t}&=& \hat{H}_{g}+\hbar\varepsilon\cos(\omega_{f}t+\phi_{f}),
\end{eqnarray}
where $\omega_{f}$ is the frequency of the fluctuation and $\phi_{f}$ is its phase. Following the steps in Appendix~\ref{app:ff}, we then obtain the filter function $F(\omega_{f})$ by calculating the infidelity (averaged over $\phi_{f}$) and dividing by $\varepsilon^{2}$. The gate infidelity in the presence of a general noise with power spectral density $P(\omega_{f})$ is then:
\begin{eqnarray}
    \mathcal{I}_{f}\simeq \int^{\infty}_{0}d\omega_{f}P(\omega_{f})F(\omega_{f}).
\end{eqnarray}
For many common noise sources, the noise power spectral density decays as $P(\omega_{f})\propto \omega_{f}^{-\gamma}$ \cite{brownnutt_2015}. Therefore, a signature of a noise-insensitive gate is that it minimizes the value of $F(\omega_{f})$ for small values of  $\omega_{f}$.

\begin{figure}[h]
\includegraphics[width=\columnwidth]{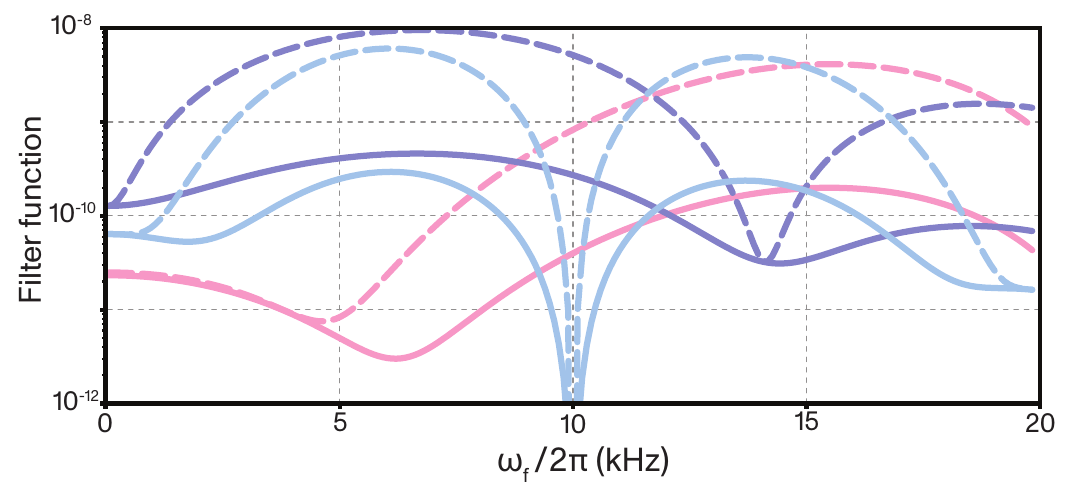}
\caption{Filter function comparison for smooth gate (pink bottom left), Walsh-3 gate (blue middle left), and Walsh-1 gate (purple top left). Each gate assumes a gradient Rabi frequency of $\Omega_{g}=2\pi\times 5~$kHz. The parameters are chosen such that the smooth and Walsh-3 gates have the same $t_{g}$. Results are for average phonon numbers $\bar{n}=0$ (solid) and $\bar{n}=10$ (dashed) in the gate mode.}
\label{fig:smooth_ff}
\end{figure}

Fig.~\ref{fig:smooth_ff} compares the filter function of the smooth gate with Walsh-1 and Walsh-3 sequence gates. Each gate assumes $\Omega_{g}\simeq 2\pi\times 5~$kHz, and ramp parameters are chosen to ensure the smooth gate and the Walsh-3 gate sequence have the same total duration $t_{g}=200~\mu$s (for the Walsh-1 gate, $t_{g}\simeq 141~\mu$s). We show results at two temperatures, with average phonon numbers $\bar{n}=0$ and $\bar{n}=10$. 

Fig.~\ref{fig:smooth_ff} illustrates three important advantages of smooth gates over Walsh modulation. First, though it takes the same amount of time as the Walsh-3 sequence, the smooth gate is approximately $2.7\times$ less sensitive to $\omega_{f}\simeq 0$ fluctuations and static offsets. Second, this advantage \textit{increases} until $\omega_{f}\sim \Omega_{g}$\textemdash indicating the smooth gate is much less sensitive to (experimentally common) time-dependent motional noise in the kHz band. Finally, we can see the $\bar{n}=0$ and $\bar{n}=10$ filter functions diverge at a much larger $\omega_{f}$ for the smooth gate, indicating its ability to suppress motional errors for a general thermal state, not only for the motional ground state. In the next section, we take advantage of this general robustness to noise and temperature to implement an ultra-high fidelity smooth gate above the Doppler limit.

\section{Experimental demonstrations}\label{sec:experiment}

\subsection{Implementation}\label{sec:experimental_setup}
We implement the smooth gate in a cryogenic ion trap setup similar to that described in \cite{loschnauer_2024}. Qubits are encoded in the $4S_{1/2}$ Zeeman sublevels of $\mathrm{^{40}Ca^+}$ ions, each with a qubit frequency of $\omega_0 \approx 2 \pi \times 240$~MHz. Each experimental shot starts with Doppler cooling via a $397$~nm laser beam $13$~MHz red-detuned from the $4S_{1/2} \leftrightarrow 4P_{1/2}$ transition and an $866$~nm beam for repumping population from $3D_{3/2}$. After cooling, the ions are optically pumped into the $\ket{\uparrow} = 4S_{1/2}\ket{m=+1/2}$ state, followed by running a sequence of two-qubit gates, and reading out the final state by shelving $\ket{\downarrow}$ to $5D_{5/2}$ and detecting fluorescence using a CMOS camera. The two-qubit gates are performed on the in-plane radial rocking motional mode of a two-ion crystal at $\omega_m = 2 \pi \times 3.3$~MHz, using a bichromatic oscillating current at frequencies $\omega_0 \pm (\omega_m + \delta(t))$ which implements a M\o lmer--S\o rensen (MS) drive. An additional low-power carrier current at $\omega_0$ provides dynamical decoupling from errors due to qubit frequency fluctuations \cite{harty_2016}. \\

At the start of each smooth gate, the amplitudes of each of the two gate tones are simultaneously ramped from 0 to $\approx 0.7$~A over $\tau_g = 5$~$\mu$s following a sin$^2$ ramp function, at a fixed detuning of $\delta(0) = \delta_\text{max} = 2 \pi \times -400$~kHz. This amplitude results in a gate Rabi frequency $\Omega_g \approx 2 \pi \times 6$~kHz. Next, the carrier tone is ramped up in amplitude over $0.5$~$\mu$s to reach a carrier Rabi frequency of $\approx 2\pi \times 80$~kHz. Subsequently, the gate tones are ramped in frequency over $\tau_d = 100$~$\mu$s to a minimum detuning $\delta_\text{min} = 2 \pi \times -21.7$~kHz following the ramp function described in Eq.~(\ref{eq:det_ramp_func}) with $j = 3$. The gate tones are held at constant amplitude and detuning $\delta_\text{min}$ for a duration $t_c = 15.8$~$\mu$s. Halfway through this duration, the phase of the carrier tone is inverted to ensure minimal residual carrier rotation by the end of the gate. Finally, the frequency ramp is reversed such that the two tones return to a detuning of $\delta(t_g) = \delta_\text{max} =  2 \pi \times -400$~kHz, and the carrier and gate tones are ramped down in amplitude over $0.5$~$\mu$s and $5$~$\mu$s respectively. The total gate duration is $t_g = 225.8$~$\mu$s. \\

\subsection{Gate calibration}\label{sec:calibration}

\begin{figure}[h]
\centering
\includegraphics[width=\columnwidth]{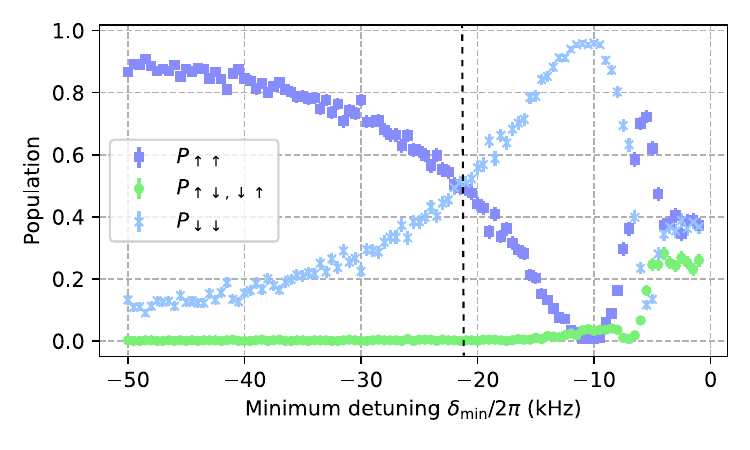}
\caption{Measured populations of $\ket{\uparrow\uparrow}$, $\ket{\downarrow\downarrow}$, and $\ket{\uparrow\downarrow}$ and $\ket{\downarrow\uparrow}$, versus minimum gate detuning $\delta_\text{min}$, for the $t_{g}\simeq 226~\mu$s pulse sequence described in the text. The dashed vertical line shows the value of $\delta_\text{min}$ where $P_{\uparrow\uparrow} \approx P_{\downarrow\downarrow} \approx 0.5$, corresponding to $\theta_g \approx \pi/2$.
}
\label{fig:gate_angle}
\end{figure}

\begin{figure*}[t]
\centering
\subfloat{\includegraphics[width=0.5\textwidth]{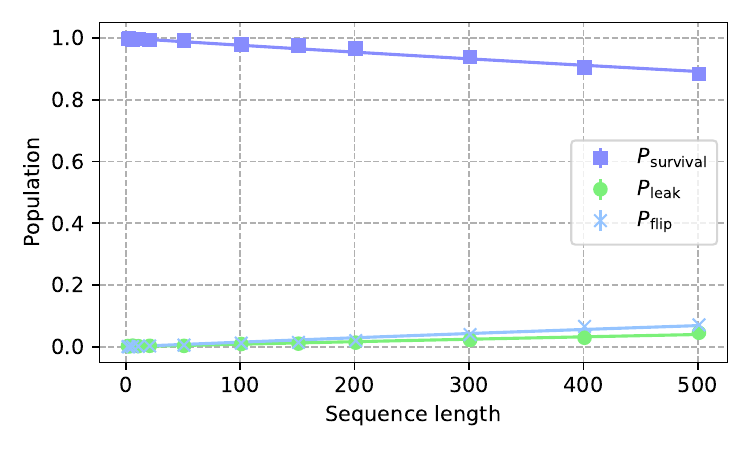}\label{fig:slrb_doppler_500_populations}}
\hfill
\subfloat{\includegraphics[width=0.5\textwidth]{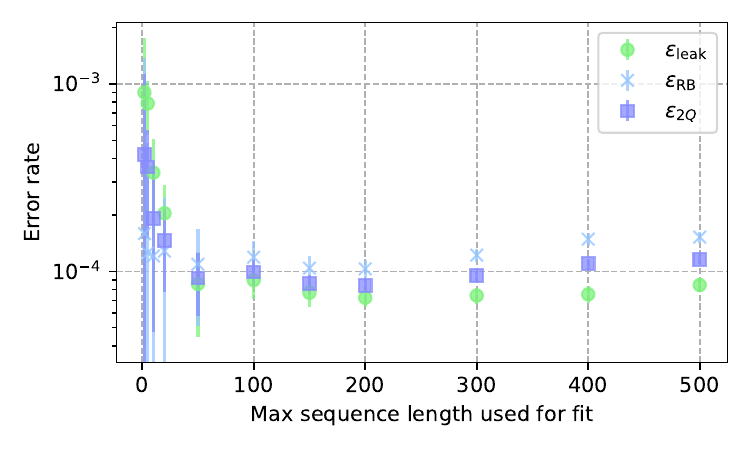}\label{fig:slrb_doppler_500_markovianity}}
\caption{SLERB of smooth gates with Doppler-cooled ions. (Left) Populations $P_{\text{survival}}$, $P_{\text{flip}}$, and $P_{\mathrm{leak}}$ after $N=2-500$ Cliffords. Each data point consists of 100 shots each of 50 random sequences of lengths \{2, 5, 10, 20, 50, 150, 300, 400, 500\}, and of 100 random sequences of lengths \{1, 100, 200\}. (Right) The leakage rate, SU(2) error rate, and inferred two-qubit gate error, extracted by fitting decay curves to the populations in the left plot. Truncating the data at different maximum sequence lengths allows characterization of non-Markovianity at the $\approx 3 \times 10^{-5}$ level. In all analyses, we assumed state-preparation and measurement (SPAM) errors to be negligible, i.e., no y-offset for the fit. Note that an extra rotation is compiled into the inverting Clifford to randomize the expected final state between $\ket{\downarrow\downarrow}$ and $\ket{\uparrow\uparrow}$, providing first-order insensitivity to any asymmetry in SPAM errors between the two states (Pauli randomization). The error bars are the $68\%$ confidence intervals extracted using non-parametric bootstrapping with 10,000 resamples.}
\label{fig:slrb_doppler_500}
\end{figure*}

To calibrate the smooth gate, it suffices to find a value of $\delta_{min}$ that corresponds to $\theta_g = \pi/2$, while simultaneously being far enough from the mode frequency to satisfy the AESE condition. To that end, we prepare the ions in $\ket{\uparrow \uparrow}$, apply a single smooth gate pulse with variable $\delta_{min}$, and measure in the computational basis. The results are shown in Fig.~\ref{fig:gate_angle}. We find that, for $|\delta_{min}|\gtrsim  2\pi\times 15~$kHz, the population in $\ket{\uparrow\downarrow}$ and $\ket{\downarrow\uparrow}$ is highly suppressed, indicating successful AESE. The maximally entangling gate $\theta_g=\pi/2$ corresponds to the point where $P_{\uparrow\uparrow} = P_{\downarrow\downarrow} = 0.5$, which is at $\delta_\text{min} = 2 \pi \times -21.7$~kHz. As discussed earlier, this single-parameter calibration makes the smooth gate very practical compared to DESE approaches, where optimizing $\theta_g$ requires a two-parameter search involving $\Omega_g$ and $\delta_g$.

\subsection{Subspace leakage randomized benchmarking at Doppler temperature}\label{sec:slerb}

We assess the performance of the smooth gate using Subspace Leakage Error Randomized Benchmarking (SLERB) \cite{benchmarking}. In brief, SLERB estimates the two-qubit gate error through a sequence of Clifford operations on the SU(2) subspace spanned by $\ket{\uparrow\uparrow}$ and $\ket{\downarrow\downarrow}$, with each Clifford built out of two-qubit gates acting in different spin bases (achieved by adjusting the phases of the bichromatic currents). Each SLERB sequence comprises only two-qubit gates, allowing two-qubit gate errors to be directly extracted from the sequence decay rate, without the need to estimate and correct for the contribution of imperfect single-qubit rotations.

In each SLERB sequence, qubits are prepared in $\ket{\uparrow\uparrow}$. Afterwards, we apply a sequence of $N$ Cliffords (each composed of, on average, 2.17 two-qubit gates), followed by measurement in the computational basis. We denote the probability of `surviving' in the correct state (which is $\ket{\uparrow\uparrow}$, unless an extra rotation was compiled into the inverting Clifford to implement Pauli randomization, in which case the correct state is $\ket{\downarrow\downarrow}$), `flipping' to other symmetric state (i.e. $\ket{\downarrow\downarrow}$ if the survival state was $\ket{\uparrow\uparrow}$), and `leaking' to the $\{\ket{\uparrow \downarrow},\ket{\downarrow\uparrow}\}$ subspace as $P_{\mathrm{survival}}$, $P_{\mathrm{flip}}$, and $P_{\mathrm{leak}}$, respectively. These probabilities are fitted assuming the functional form in Eq.(25) in \cite{benchmarking} to obtain the `leakage error rate' $\varepsilon_\text{leak}$ and the `SU(2) error rate' $\varepsilon_{\mathrm{RB}}$ (note that, for $N \times (\varepsilon_\text{leak} + \varepsilon_\text{flip}) \ll 1$, these reduce to $P_{\mathrm{leak/flip}} \approx N \times \varepsilon_\text{leak/flip}$). Following Eq.~(32) in \cite{benchmarking}, the error per gate is then given by:
\begin{equation}
    \varepsilon_{2q} = \left ( \frac{6}{5}\varepsilon_{\mathrm{RB}} + \frac{4}{5}\varepsilon_\text{leak}\right ) \times \frac{6}{13}.
\end{equation}
As discussed in Ref.~\cite{benchmarking}, the magnitudes of $\varepsilon_{\mathrm{RB}}$ and $\varepsilon_\text{leak}$ provide insight into the origin of gate errors, with geometric phase errors contributing solely to $\varepsilon_{\mathrm{RB}}$, and spin-motion entanglement contributing to both. Thus, SLERB gives us the ability to directly estimate not just the fidelity of gates, but also the effectiveness of AESE.

Fig.~\ref{fig:slrb_doppler_500} (left) shows SLERB population measurements with up to $N=500$ sequential Cliffords (1083 gates). Fitting the whole dataset, we find an average smooth two-qubit gate error $\varepsilon_{2q} = 1.16(6) \times 10^{-4}$ on ions prepared near the Doppler limit (note no mid-circuit cooling was used to maintain temperature throughout the SLERB sequence). We further find that the SU(2) error $\varepsilon_{\mathrm{RB}} \approx 1.5 \times 10^{-4}$ dominates over the leakage error of $\varepsilon_\text{leak} \approx 8 \times 10^{-5}$, indicating that the smooth gate effectively suppresses spin-motion entanglement. We attribute most of the residual gate errors to drifts of the magnetic-field-sensitive qubit frequency across the gate sequence.

SLERB allows us to get further insight into the time-dependence of gate errors by truncating the maximum sequence length used for fitting the decay curve. The results, shown in Fig.~\ref{fig:slrb_doppler_500} (right), indicate a small but statistically significant amount of non-Markovianity in the gate sequence. For sequences of up to $N=200$ Cliffords (432 two-qubit gates), we find a gate error as low as $8.4(7) \times 10^{-5}$, which then increases by $\approx 3 \times 10^{-5}$ for longer sequences. Note that we only observe memory effects in $\varepsilon_{\mathrm{RB}}$, with spin-motion entanglement errors remaining at $\varepsilon_\text{leak} \approx 8 \times 10^{-5}$ regardless of sequence length. While detailed characterization of gate memory effects is beyond the scope of this work, the remarkably low level of non-Markovianity achieved here \emph{despite} the use of magnetic-field-sensitive qubits demonstrates the extreme promise of laser-free smooth gates for fault-tolerant quantum computing.

\subsection{Robustness to detuning offsets}\label{sec:mot_demo}

\begin{figure}[h]
\centering
\includegraphics[width=\columnwidth]{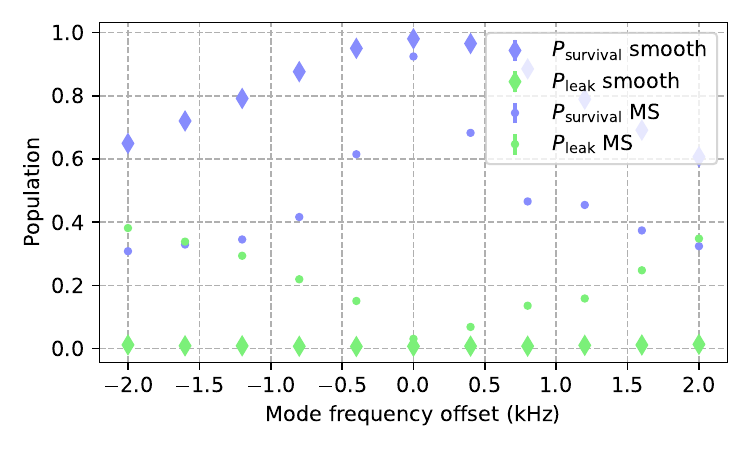}
\caption{Average survival and leakage probability after 50 randomizations of SLERB sequences of length $N=100$, at the Doppler temperature. The results for smooth gates are shown as diamonds, and compared to Walsh-1 MS gates (dots). A static error of the gate mode frequency was simulated by adding an offset to $\delta_g$ in the case of the MS gate, or to $\delta(t)$ for the smooth gate.}
\label{fig:detuning_scans}
\end{figure}

To verify the robustness of the smooth gate to motional frequency errors, we look at outcomes of the SLERB sequences with $N=100$ Cliffords in the presence of static mode frequency offsets over a range of $\pm 2~$kHz. Following Sec.~\ref{sec:smooth_robust}, we also compare it to the Walsh-1 MS gate of approximately the same gate Rabi frequency (total gate duration of $124~\mu$s). 

The results, shown in Fig.~\ref{fig:detuning_scans}, illustrate two important properties of the smooth gate. Firstly, both $P_\mathrm{survival}$ and $P_{\mathrm{leak}}$ (and thus also $\varepsilon_{2q}$), are much less sensitive to detuning offsets for the smooth gate relative to the Walsh-1 gate. Secondly, as anticipated in Sec.~\ref{sec:smooth_robust}, this decreased sensitivity comes primarily from a reduction in spin-motion entanglement errors; this is because such errors manifest themselves as $\varepsilon_\text{leak}$ and, in turn, $P_{\mathrm{leak}}$. Therefore, these results indicate that, for smooth gates, mode frequency offsets indeed mainly lead to (temperature-insensitive) errors in $\theta_{g}$, and do not cause significant (temperature-sensitive) spin-motion entanglement errors.

\subsection{Benchmarking at different starting temperatures}\label{sec:doppler_demo}
To directly measure how the smooth gate error depends on gate mode occupation, we prepare the ion pair at different initial temperatures. This is achieved either by performing sideband cooling directly after Doppler cooling (reducing temperature of all motional modes below the Doppler limit), or by deliberately spoiling Doppler cooling by tuning the frequency of the $397$~nm laser closer to resonance (increasing temperature above the Doppler limit). We characterize the initial temperature of the gate mode in each case via a separate set of experiments, using a $729$~nm sideband probe pulse of varying duration and fitting the resulting two-ion populations $P_{\uparrow\uparrow}$, $P_{\downarrow\downarrow}$ and $P_{\uparrow\downarrow\\,\downarrow\uparrow}$ as a function of the pulse duration, assuming an initial thermal distribution of motional states \cite{home_thesis}.

\begin{figure}[h]
\includegraphics[width=0.99\columnwidth]{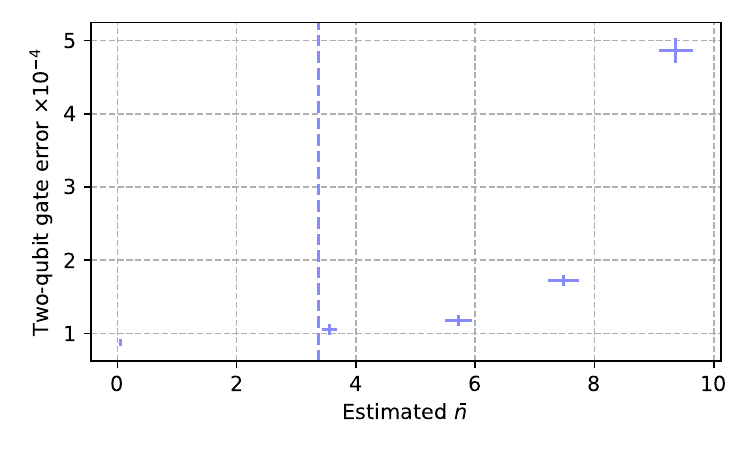}
\caption{The two-qubit smooth gate error rate vs initial gate mode temperature. Each measurement represents 
a SLERB experiment consisting of 100 shots each of 50 random sequences of lengths \{1, 2, 5, 10, 20, 50, 100, 150, 200\}, with randomization of the final state. The vertical dashed line indicates the theoretical Doppler limit. The error bars are the $68\%$ confidence intervals extracted using non-parametric bootstrapping with 10,000 resamples.
}
\label{fig:slrb_vs_temp}
\end{figure}

The measurement results of gate error as a function of initial temperature are shown in Figure \ref{fig:slrb_vs_temp}. The measurements are performed at five different temperatures: near the motional ground state ($\bar{n} = 0.053(2)$), and then at four temperatures starting near the Doppler limit ($\bar{n} = 3.5(1)$) and up to $\bar{n}=9.4(3)$. At the lowest temperature, we record an error of $\varepsilon_{2q} = 8.8(5) \times 10^{-5}$. The error remains at $\varepsilon_{2q} \lesssim 5 \times 10^{-4}$ over the full range of temperatures probed, indicating capability for high-fidelity quantum operations even far above the Doppler limit. Note that, in this dataset, the gate calibration was only performed once with ground-state-cooled ions: optimizing directly for a specific $\bar{n}$ might allow for additional performance improvements.

\section{Conclusions}\label{sec:architecture}
To summarize, we have introduced the smooth gate: a new adiabatic method for performing entangling gates in trapped-ion systems. We have validated its performance on a pair of electronically controlled $\mathrm{^{40}Ca^+}$ ions, achieving record-high fidelity of $>0.9999$ while remaining above the Doppler limit. The ability to perform quantum logic with error rates $\sim 10^{-4}$ has significant implications for trapped-ion quantum computing: it opens the door to NISQ applications with tens of thousands of gates per circuit \cite{Clinton2024}, as well as to low-overhead QEC, e.g., using qLDPC codes \cite{Bravyi2024, Malcolm2025}. However, the appeal of the smooth gate for large-scale trapped-ion QCs goes far beyond the headline fidelity number. 

First, the smooth gate significantly simplifies the implementation of QC architectures based on global qubit drives, such as discussed in \cite{leibfried_2007, malinowski_2023}. Briefly: a smooth gate can be performed in a single time step on all qubit pairs coupled to the same gradient source, with all zone-to-zone variations (e.g., stray fields, fabrication defects, etc.) shimmed by locally fine-tuning the gate mode frequency using local DC electrodes. Deliberate mode frequency offsets can be introduced through the same electrodes to locally turn off the interactions (gate addressing), or to fine-tune the entangling angle on a qubit-by-qubit basis (fractional-angle gates). In large-scale QCs, those DC electrodes can be multiplexed on chip \cite{malinowski_2023}, alleviating the `QC wiring bottleneck'.

The second, and arguably the most impactful, benefit of smooth gates is that they open the door to electronic trapped-ion QCs operating fully above the Doppler limit. This is highly significant, as ground-state cooling is the primary driver of total circuit runtime in modern QCCD architectures (e.g. $\sim 25-68\%$ in \cite{moses_2023}). Furthermore, the smooth gate's relaxed thermal occupation requirements lead to relaxed transport-induced heating and mode excitation requirements -- in turn allowing for faster ion movement across the QCCD device. As transport and cooling together account for $\sim 98-99\%$ of the duration of a typical QCCD circuit \cite{pino_2021,moses_2023}, our approach holds clear potential for over an order of magnitude speedup of circuit execution. Finally, the ability to eliminate sub-Doppler cooling significantly simplifies the device design and engineering by removing complexity. For example, ground-state cooling typically requires significantly more laser power compared to Doppler cooling; it also creates extra constraints on laser beam orientations, polarizations, or spatial mode structure \cite{Corsetti2025, Xing2025}. Thus, eliminating ground-state cooling altogether simplifies the optical engineering of large-scale trapped-ion QCs. All in all, smooth gates unlock trapped-ion QC architectures that are simultaneously better, faster, and simpler compared to alternatives. 

\section*{Acknowledgments}
We thank the entire combined team at IonQ for their contributions to this work.

\bibliography{biblio}
\newpage

\appendix

\section{Rotating frame transformation}\label{app:rotating_frame}
In the literature, phase-space trajectories are typically shown in the rotating frame with respect to the motion, which we can obtain by transforming Eq.~(\ref{eq:geo_general_lab}) according to:
\begin{eqnarray}\label{eq:mot_int_u}
    \hat{U}_{I}&=&\exp\Big(-i\hat{a}^{\dagger}\hat{a}\int^{t}_{0}dt^{\prime}\delta[t^{\prime}] \Big),
\end{eqnarray}
mapping the position operator onto:
\begin{eqnarray}
    \hat{x}&\rightarrow & \hat{x}\cos(\varepsilon[t])+\hat{p}\sin(\varepsilon[t]),
\end{eqnarray}
where $\varepsilon(t)\equiv \int^{t}_{0}dt^{\prime}\delta(t^{\prime})$. When the gate is adiabatic and following a trajectory near the interaction-frame position axis, the same operation will spiral about the phase-space origin when described in the rotating frame.
\section{Infidelity due to mode frequency fluctuations}\label{app:mode_freq_inf}

While there are other kinds of geometric phase gates \cite{sutherland_2021_2}, we will define a two-qubit geometric phase gate as one that 1) entangles two ions via shared motional modes that 2) can be satisfactorily approximated as:
\begin{eqnarray}\label{eq:gate_ham_rot}
    \hat{H}_{g}&=& \frac{\hbar\Omega_{g}f(t)}{2}\hat{S}_{\alpha}\Big(\hat{a}^{\dagger}e^{i\phi[t]} + \hat{a}e^{-i\phi[t]}e^{-i\phi(t)} \Big),
\end{eqnarray}
in some frame of reference. Each scheme we consider will be for a specific combination of envelope functions $-1\leq f(t)\leq1$, two-qubit Pauli operator $\hat{S}_{\alpha}=\hat{\sigma}_{\alpha,1}\pm\hat{\sigma}_{\alpha,2}$ and time varying phase $\phi(t)$. The time-propagator for this system can always be exactly described by the Magnus expansion \cite{magnus_1954,roos_2008}:
\begin{eqnarray}
    \hat{U}_{g}(t)&=&\exp\Big(\frac{-i}{\hbar}\int^{t}_{0}dt^{\prime}\hat{H}_{g}(t^{\prime}) \nonumber \\
    &&-\frac{1}{2\hbar^{2}}\int^{t}_{0}\int^{t^{\prime}_{0}}dt^{\prime}dt^{\prime\prime}\Big[\hat{H}_{g}(t^{\prime}),\hat{H}_{g}(t^{\prime\prime})\Big] \Big) \nonumber \\
\end{eqnarray}
which, for any $\{f(t),\phi(t)\}$, always takes the form:
\begin{eqnarray}\label{eq:simple_mag}
\hat{U}_{g}(t)\!\!&=&\!\! \exp\Big\{(\gamma[t]\hat{a}^{\dagger}-\gamma^{*}[t]\hat{a})\hat{S}_{\alpha}\Big\}\exp\Big\{-\frac{i\theta[t]}{2}\hat{S}_{\alpha}^{2} \Big\}.
\end{eqnarray}
where $\gamma(t)$ the displacement associated with the spin-motion entanglement, and $\theta(t)$ is the gate angle at time $t$. At the gate time $t_{g}$, the time propagator would ideally be:
\begin{eqnarray}
    \hat{U}_{g}(t_{g})&=&\exp\Big\{-\frac{i\theta_{g}}{2}\hat{S}_{\alpha}^{2} \Big\},
\end{eqnarray}
where $\theta_{g}$ is the gate entanglement angle and $\gamma(t_{g})=0$, i.e. there is no residual spin-motion entanglement. \\

If the modes fluctuate during the gate, this will give an error Hamiltonian:
\begin{eqnarray}
    \hat{H}_{e}(t)&=&\hbar\varepsilon(t)\hat{a}^{\dagger}\hat{a},
\end{eqnarray}
making the total Hamiltonian:
\begin{eqnarray}
    \hat{H}_{t}&=& \hat{H}_{g} + \hat{H}_{e},
\end{eqnarray}
which we can then transform into the interaction picture with respect to $\hat{H}_{e}$, mapping the error onto a new phase function $\phi_{t}(t)=\phi_{g}(t)+\int^{t}_{0}dt^{\prime}\varepsilon(t^{\prime})$. Since the form of Eq.~(\ref{eq:simple_mag}) is true for any real $\phi(t)$, we know that $\hat{U}_{t}(t_{g})$ can be rewritten in factored form $\hat{U}_{g}\hat{U}_{e}$:
\begin{eqnarray}
     \hat{U}_{e}&= &\exp\Big\{(\gamma_{t}\hat{a}^{\dagger}-\gamma_{t}^{*}\hat{a})\hat{S}_{\alpha}\Big\}\exp\Big\{\frac{i\varepsilon\theta}{2}\hat{S}_{\alpha}^{2} \Big\}
\end{eqnarray}
where $\gamma_{t}$ is the displacement associated with the residual spin-motion entanglement at $t_{g}$, and $\varepsilon\theta\equiv \theta_{t}-\theta_{g}$\textemdash which is possible due to the fact that $[\hat{U}_{g},\hat{U}_{t}]=0$. We can immediately plug this into our equation for fidelity: 
\begin{eqnarray}\label{eq:fidelity}
\mathcal{F} &= &\sum_{n^{\prime}}|\bra{T}\bra{n^{\prime}}\hat{U}_{g}\hat{U}_{e}\ket{\psi_{0}\ket{n}}|^{2} \nonumber \\
&=&\sum_{n^{\prime}}|\bra{\psi_{0}}\bra{n^{\prime}}\hat{U}_{e}\ket{\psi_{0}\ket{n}}|^{2},
\end{eqnarray}
where $\ket{\psi_{0}}$ is the initial state of the qubits, and $\ket{T}\equiv \hat{U}_{g}\ket{\psi_{0}}$ is the target state. Assuming $\mathcal{F}\simeq 1$, we expand $\hat{U}_{e}$ in this equation, keeping only quadratic terms:
\begin{eqnarray}\label{eq:infidelity}
    \mathcal{I} = (2n+1)|\alpha_{t}|^{2}\lambda^{2}_{\hat{S}_{\alpha}} + \frac{\delta\theta^{2}}{4}\lambda^{2}_{\hat{S}_{\alpha}^{2}},
\end{eqnarray}
where the first term represents the infidelity due to residual spin-motion entanglement $\mathcal{I}_{\gamma}$, and the second term is the error in the gate angle $\mathcal{I}_{\theta}$. The dependence of each component of $\mathcal{I}$ on $\ket{\psi_{0}}$ is represented with the variance $\lambda_{\hat{S}_{\alpha}^{1(2)}}^{2}$. Knowing this about the spin-dependence of the quadratic terms in $\mathcal{I}$ will allow us to simplify our calculations when determining the effect of $\varepsilon(t)$ on gate fidelities. \\

Saying we are only interested in the high-fidelity limit is akin to saying we are interested in the perturbative effect that $\hat{H}_{e}$ has on the time evolution of $\hat{H}_{g}$. Using this fact, we can transform into the interaction picture with respect to the ideal Hamiltonian:
\begin{eqnarray}
    \tilde{H}_{e} &= & \hat{U}_{g}(t)^{\dagger}\hat{H}_{t}\hat{U}_{g} + i\hbar\dot{\hat{U}}_{g}^{\dagger}\hat{U}_{g} \nonumber \\
    &=& \hbar\varepsilon(t)\Big\{\hat{a}^{\dagger}\hat{a} + \hat{S}_{\alpha}(\gamma[t]\hat{a}^{\dagger}+\gamma^{*}[t]\hat{a}) + |\gamma[t]|^{2}\hat{S}^{2}_{\alpha}\Big\}, \nonumber \\
\end{eqnarray}
which we can immediately plug into $2^{\text{nd}}$-order time dependent perturbation theory to obtain:
\begin{eqnarray}
    \tilde{U}_{e}(t_{g})=\hat{I} -\frac{i}{\hbar}\!\!\int^{t_{g}}_{0}\!\!\!\!dt^{\prime}\tilde{H}_{e}(t^{\prime})-\frac{1}{\hbar^{2}}\!\!\int^{t_{g}}_{0}\!\!\!\int^{t^{\prime}}_{0}\!\!\!\!\!dt^{\prime}dt^{\prime\prime}\tilde{H}_{e}(t^{\prime})\hat{H}_{e}(t^{\prime\prime})\nonumber ,\\
\end{eqnarray}
which we can then plug into Eq.~(\ref{eq:fidelity}). Keeping only terms that are quadratic in our error term gives:
\begin{widetext}
\begin{eqnarray*}
    \mathcal{I} &=& 2\text{Re}\Big[\int^{t_{g}}_{0}\int^{t^{\prime}}_{0}dt^{\prime}dt^{\prime\prime}\varepsilon(t^{\prime})\varepsilon(t^{\prime\prime})\Big\{\braket{\hat{S}^{2}_{\alpha}}(2n+1)\gamma(t^{\prime})\gamma^{*}(t^{\prime\prime})  +\braket{\hat{S}_{\alpha}^4}|\gamma(t^{\prime})|^{2}|\gamma(t^{\prime\prime})|^{2}\Big\}\Big] \nonumber \\
    && - \braket{\hat{S}_{\alpha}}^{2}(2n+1)\Big|\int^{t_{g}}_{0}dt^{\prime}\varepsilon(t^{\prime})\gamma(t^{\prime})\Big|^{2} \nonumber -\braket{\hat{S}_{\alpha}^{2}}^{2}\Big|\int^{t_{g}}_{0}dt^{\prime}\varepsilon(t^{\prime})|\gamma(t^{\prime})|^2 \Big|^{2}.
\end{eqnarray*}
\end{widetext}
For dependence of this equation for $\mathcal{I}$ on the initial state to match that of Eq.~(\ref{eq:infidelity}), we know that:
\begin{eqnarray*}
\!2\text{Re}\Big[\int^{t_{g}}_{0}\!\!\!\int^{t^{\prime}}_{0}\!\!\!dt^{\prime}dt^{\prime\prime}\varepsilon(t^{\prime})\varepsilon(t^{\prime\prime})\gamma(t^{\prime})\gamma^{*}(t^{\prime\prime})\Big] \!\!\!\!&=&\!\!\!\Big|\!\int^{t_{g}}_{0}\!\!\!\!\!dt^{\prime}\varepsilon(t^{\prime})\gamma(t^{\prime})\Big|^{2} \nonumber \\
\!\!\!\!\!\!2\text{Re}\Big[\!\int^{t_{g}}_{0}\!\!\!\!\!\int^{t^{\prime}}_{0}\!\!\!\!\!dt^{\prime}dt^{\prime\prime}\!\varepsilon(t^{\prime})\varepsilon(t^{\prime\prime})|\gamma(t^{\prime})|^{2}|\gamma(t^{\prime\prime})|^{2}\Big]\!\!\!\!&=&\!\!\!\Big|\!\!\int^{t_{g}}_{0}\!\!\!\!\!dt^{\prime}\varepsilon(t^{\prime})|\gamma(t^{\prime})|^2 \Big|^{2}, \nonumber \\
\end{eqnarray*}
which simplifies the infidelity equation to: 
\begin{eqnarray}\label{eq:infidelity_integral}
    \mathcal{I} = (2n+1)&\Big|\int^{t_{g}}_{0}dt^{\prime}\varepsilon(t^{\prime})\gamma(t^{\prime}) \Big|^{2}\lambda_{\hat{S}_{\alpha}}^{2}+ \nonumber \\
    &\Big|\int^{t_{g}}_{0}dt^{\prime}\varepsilon(t^{\prime})|\gamma(t^{\prime})|^{2} \Big|^{2}\lambda_{\hat{S}_{\alpha}^{2}}^{2}, \nonumber \\
\end{eqnarray}
reducing the infidelity due to two integral equations. Up to a phase, these two integrals correspond to the residual spin-dependent displacement:
\begin{eqnarray}\label{eq:sd_disp} \alpha_{t}=\int^{t_{g}}_{0}dt^{\prime}\varepsilon(t^{\prime})\gamma(t^{\prime}),
\end{eqnarray}
and the entanglement angle error:
\begin{eqnarray}
\delta\theta&=&2\int^{t_{g}}_{0}dt^{\prime}\varepsilon(t^{\prime})|\gamma(t^{\prime})|^{2}.
\end{eqnarray}
Since we can change the sign of $\gamma$ during the gate, it is possible to implement gate protocols that minimize $\mathcal{I}_{\gamma}$ by choosing a function for $\gamma(t)$ that time-averages $\alpha_{t}\sim 0$; this should not be possible for $\mathcal{I}_{\theta}$, since we cannot change the sign of its respective integrand in the same way.

\subsection{Filter function}\label{app:ff}

Assume the error Hamiltonian takes the form:
\begin{eqnarray}
    \hat{H}_{e}(t)&=& \hbar\varepsilon_{0}\cos(\omega t +\phi)\hat{a}^{\dagger}\hat{a}.
\end{eqnarray}
We now insert this into Eq.~(\ref{eq:infidelity_integral}) and divide by $\varepsilon_{0}^{2}$:
\begin{eqnarray}\label{eq:filter_function_phi}
    S(\omega)\simeq &=& (2n+1)\Big|\int^{t_{g}}_{0}dt^{\prime}\cos(\omega t^{\prime}+\phi)\gamma(t^{\prime}) \Big|^{2}\lambda_{\hat{S}_{\alpha}}^{2} \nonumber \\
    &&+\Big|\int^{t_{g}}_{0}dt^{\prime}\cos(\omega t^{\prime}+\phi)|\gamma(t^{\prime})|^{2} \Big|^{2}\lambda_{\hat{S}_{\alpha}^{2}}^{2} \nonumber \\
\end{eqnarray}
giving a shift-independent `filter-function', recasting each integral as Fourier transforms of $\gamma(t)$ and $|\gamma(t)|^{2}$. Averaging over $\phi$ gives:
\begin{eqnarray}\label{eq:filter_function_avg}
    S(\omega)& = &\frac{1}{2}\sum_{\phi}\Big\{(2n+1)\Big|\int^{t_{g}}_{0} dt^{\prime}\!\!\cos(\omega t^{\prime}+\phi)\gamma(t^{\prime}) \Big|^{2}\nonumber \\
&&+\Big|\int^{t_{g}}_{0}dt^{\prime}\cos(\omega t^{\prime}+\phi)|\gamma(t^{\prime})|^{2} \Big|^{2}\Big\}
\end{eqnarray}
where the sum is over $\phi\in\{0,-\pi/2\}$; in other words we simply average over the cases when $\varepsilon(t)$ follows a $\cos$ and $\sin$ function:
\begin{eqnarray}\label{eq:alpha_phi_avg}
    |\alpha_{\phi}|^{2} &= & \frac{1}{2}\Big(|\alpha_{t,\cos}|^{2}+|\alpha_{t,\sin}|^{2} \Big) \nonumber \\
    \delta \theta^{2}_{\phi} &=& \frac{1}{2}\Big(\theta^{2}_{\text{cos}}+\theta^{2}_{\text{sin}} \Big).
\end{eqnarray}
In the limit where $\mathcal{I}_{\alpha}$ is small enough such that the assumption that errors are Markovian and quadratic is valid, we can integrate $S(\omega)$ against the (independently measured) power spectral density of mode frequency fluctuations:
\begin{eqnarray}
    \mathcal{I}=\int^{\infty}_{0}d\omega P(\omega)S(\omega),
\end{eqnarray}
giving an estimate of the total error due to mode frequency fluctuations\textemdash both static and time-dependent. For the specific examples below, we will discuss each scheme's robustness to motional frequency fluctuations in terms of $S(\omega)$ for that sequence, as it provides information about a scheme's robustness to static \textit{and} time-dependent mode frequency fluctuations. 

\subsubsection{Filter function for Walsh sequenced diabatic gates}\label{app:ff_walsh}

\begin{figure}[ht]
\centering
\includegraphics[width=0.95\columnwidth]{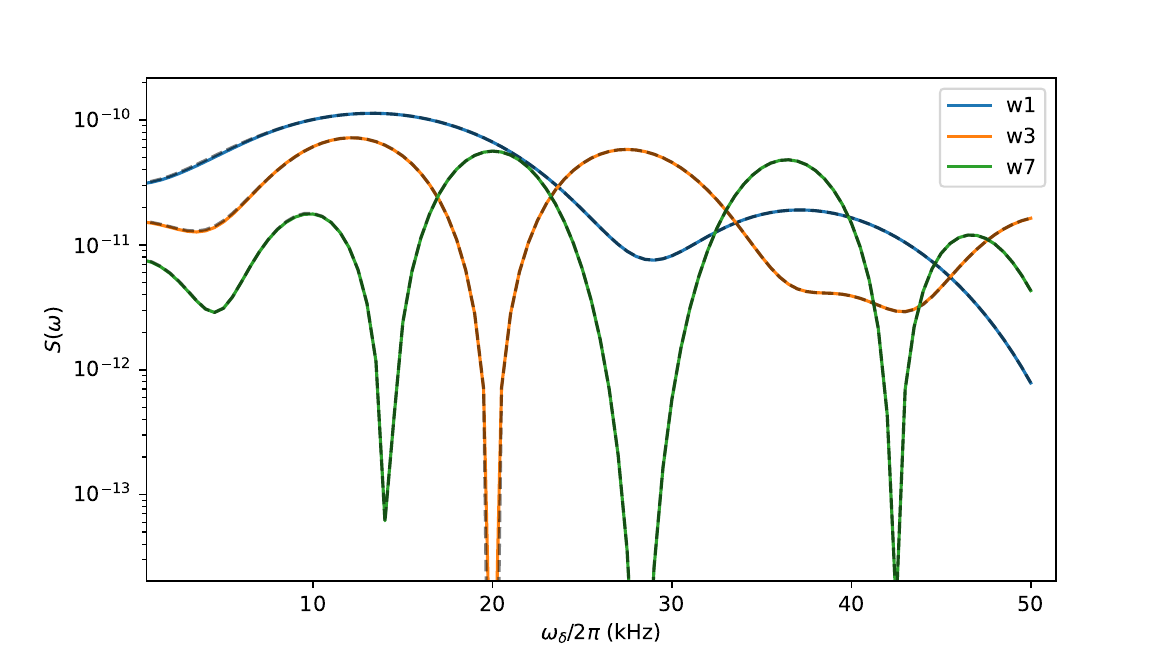}
\caption{Comparison between the analytical expression for Walsh-K filter function in Eq.~(\ref{eq:walsh_ff}) to the direct numerical integration of Eq.~(\ref{eq:gate_ham_rot}).
}
\label{fig:walsh_ff_compare}
\end{figure}

Walsh gates are a set of gate schemes that, at integer multiples of $2\pi/\delta$, set $\gamma\rightarrow -\gamma$, averaging the effect of static motional frequency shifts to zero for increasing powers of $\delta$ \cite{hayes_2012}. For any sequence higher-order than a Walsh-0 (i.e., a single loop gate with no echo), this causes $S(0)\rightarrow 0$, signifying an increased resilience to mode frequency fluctuations that are slow relative to an individual loop. To understand how non-static shifts map onto $\mathcal{I}_{\gamma}$ for Walsh gates, we employ Eq.~(\ref{eq:sd_disp}), which, while neglecting the higher-order contributions to static shifts, provides an accurate representation of the behaviour due to time-dependent shifts. To calculate $\alpha_{t}$ for a $K-1$ Walsh sequence, we can divide Eq.~(\ref{eq:sd_disp}) into $K$ integrals:
\begin{eqnarray}
    \alpha_{t}=\sum_{m=0}^{K-1}\int^{t_{m+1}}_{t_{m}}dt^{\prime}\varepsilon(t^{\prime})\gamma_{m}(t^{\prime}),
\end{eqnarray}
where $t_{m}\equiv 2\pi m/\delta$, and:
\begin{eqnarray}
    \gamma_{m}(t) &=& W_{m}^{K-1}\frac{i\Omega_{g}}{2}\int_{t_{m}}^{t}dt^{\prime}e^{-i\delta t^{\prime}} \nonumber \\
    &=& W_{m}^{K-1}\frac{\Omega_{g}}{2\delta}(e^{-i\delta t^{\prime}}-1),
\end{eqnarray}
where $W_{m}^{K-1}=\pm 1$ is the value of the $K-1$ Walsh function for loop $m$, and $t_{m}\equiv 2\pi m/\delta$. Since Walsh sequences have no effect on $\mathcal{I}_{\delta \theta}$, it is not necessary to do this, and we can simply integrate from $0$  to $t_{g}$. Thus, to determine $S_{W,K}$, we calculate four integrals. The first two give us the residual spin-dependent displacements:
\begin{widetext}
\begin{eqnarray}
    \frac{\alpha_{m}^{\cos}}{\varepsilon_{0}} &= W_{m}^{K-1}\frac{\Omega_{g}}{2\delta}\int^{t_{m+1}}_{t_{m}}dt^{\prime}\cos(\omega t^{\prime})(e^{-i\delta t^{\prime}}-1) = W_{m}^{K-1}\frac{\Omega_{g}\sin[\frac{\pi\omega}{\delta}]}{\omega(\omega^{2}-\delta^{2})} \Big\{\!\delta\cos\!\Big[\frac{(2m\!\!+\!\!1)\pi\omega}{\delta}\Big]\!\!+\!i\omega\sin\!\Big[\frac{(2m\!\!+\!\!1)\pi\omega}{\delta}\Big] \Big\}, \nonumber \\
    \frac{\alpha_{m}^{\sin}}{\varepsilon_{0}} &= W_{m}^{K-1}\frac{\Omega_{g}}{2\delta}\int^{t_{m+1}}_{t_{m}}dt^{\prime}\sin(\omega t^{\prime})(e^{-i\delta t^{\prime}}-1) = W_{m}^{K-1}\frac{\Omega_{g}\sin[\frac{\pi\omega}{\delta}]}{\omega(\omega^{2}-\delta^{2})}\Big\{\!\delta\sin\!\Big[\frac{(2m\!\!+\!\!1)\pi\omega}{\delta}\Big]\!\!-\!i\omega\cos\!\Big[\frac{(2m\!\!+\!\!1)\pi\omega}{\delta}\Big] \Big\}, \nonumber
\end{eqnarray}
\end{widetext}
which we can use to obtain the average spin-dependent displacement when averaged over all values of $\phi$:
\begin{eqnarray}
    |\alpha_{t,K}|^{2} = \frac{1}{2}\Big(\Big|\sum_{m}\alpha_{m,\cos} \Big|^{2}+\Big|\sum_{m}\alpha_{m,\sin} \Big|^{2} \Big).
\end{eqnarray}
Since Walsh sequences have no effect on $\mathcal{I}_{\delta \theta}$, it is not necessary to add each loop separately. We can therefore integrate from $0$  to $t_{g}$ to calculate the remaining two integrals:
\begin{eqnarray*}
    \frac{\delta \theta_{\cos}^{2}}{4\varepsilon_{0}^{2}} &=& \frac{\Omega_{g}^{2}}{\delta^{2}}\Big[\int^{t_{g}}_{0}dt^{\prime}\cos(\omega t^{\prime})\sin^{2}(\delta t^{\prime}/2)\Big]^{2}\lambda^{2}_{\hat{S}^{2}_{\alpha}} \nonumber \\
    &=& \frac{\Omega_{g}^{4}}{4[4\Omega_{g}^{2}\omega K - \omega^{3}]^{2}}\sin^{2}\Big(\frac{\pi\omega K^{1/2}}{\Omega_{g}}\Big)\lambda^{2}_{\hat{S}^{2}_{\alpha}},
\end{eqnarray*}
and
\begin{eqnarray*}
    \frac{\delta \theta_{\sin}^{2}}{4} &=& \frac{\Omega_{g}^{2}\varepsilon_{0}^{2}}{\delta^{2}}\Big[\int^{t_{g}}_{0}dt^{\prime}\sin(\omega t^{\prime})\sin^{2}(\delta t^{\prime}/2)\Big]^{2}\lambda^{2}_{\hat{S}^{2}_{\alpha}} \nonumber \\
    &=& \frac{\Omega_{g}^{4}\varepsilon_{0}^{2}}{[4\Omega_{g}^{2}\omega K - \omega^{3}]^{2}}\sin^{4}\Big(\frac{\pi\omega K^{1/2}}{2\Omega_{g}}\Big)\lambda^{2}_{\hat{S}^{2}_{\alpha}}.
\end{eqnarray*}
We can calculate $\mathcal{I}$ by plugging these values into Eq.~(\ref{eq:infidelity}), diving that by $\varepsilon_{0}^{2}$, we get the filter function for a Walsh $K-1$ sequence:
\begin{eqnarray}\label{eq:walsh_ff}
    S(\omega) &= (2n+1)|\alpha_{t,K}^{\prime}|^{2}\lambda^{2}_{\hat{S}_{\alpha}}+ \nonumber \\ &\frac{\Omega_{g}^{4}(4\sin^{4}[\frac{k\omega}{2}] + \sin^{2}[k\omega])}{4[4\Omega_{g}^{2}\omega K - \omega^{3}]^{2}}\lambda^{2}_{\hat{S}^{2}_{\alpha}}, \nonumber \\
\end{eqnarray}
where $\alpha^{\prime}_{j}\equiv \alpha_{j}/\varepsilon_{0}$, and $k\equiv \pi K^{1/2}/\Omega_{g}$. In the limit $\omega \gg 2\Omega_{g}K^{1/2}$, $S_{\gamma}(\omega)\propto \omega^{-4}$ and $S_{\delta\theta}(\omega)\propto \omega^{-6}$; spin-motion entanglement errors should dominate for high-frequency noise. In Fig.~\ref{fig:walsh_ff_compare}, we compare Eq.~(\ref{eq:walsh_ff}) to the direct numerical integration of Eq.~(\ref{eq:gate_ham_rot}). The two calculations should converge in the high-fidelity limit.

\subsubsection{Filter function for adiabatic gates}\label{app:ff_smooth}

As discussed in the main text, Eq.~(\ref{eq:first_order_adiabatic_u}) describes the propagator for the full system; in the adiabatic limit ($\dot{\alpha}(t)/\delta(t)\ll 1$), $\hat{D}_{1}$ encapsulates the system's spin-motion entanglement at all times. We can rewrite this operator in the (more commonly used) rotating frame with respect to the gate mode detuning using the transformation:
\begin{eqnarray}
    \hat{U}_{\eta}&\equiv & \exp\Big(-i\int^{t}_{0}dt^{\prime}\delta[t^{\prime}]\hat{a}^{\dagger}\hat{a} \Big) \nonumber \\
    &\equiv &e^{-i\eta(t)\hat{a}^{\dagger}\hat{a}},
\end{eqnarray}
which gives:
\begin{eqnarray}
    \tilde{D}_{1}\simeq \exp\Big(\mp\frac{\alpha[t]}{2\delta[t]}\hat{S}_{\alpha}\Big[\hat{a}^{\dagger}e^{i\eta(t)}-\hat{a}e^{-i\eta(t)} \Big] \Big),
\end{eqnarray}
where we have reintroduced the spin-dependence of the operator. This tells us:
\begin{eqnarray}
    \gamma(t)&=&\mp \frac{\alpha(t)}{\delta(t)}e^{i\eta(t)}.
\end{eqnarray}
We can now plug this into Eq.~(\ref{eq:infidelity_integral}) and evaluate the spin-dependent displacement terms:
\begin{eqnarray}
    \alpha_{t,\cos} &=& \frac{1}{2}\int^{t_{g}}_{0}dt^{\prime}\cos(\omega t^{\prime})\frac{\Omega(t^{\prime})}{\delta(t^{\prime})}e^{i\eta(t^{\prime})} \nonumber \\
    \alpha_{t,\sin} &=& \frac{1}{2}\int^{t_{g}}_{0}dt^{\prime}\sin(\omega t^{\prime})\frac{\Omega(t^{\prime})}{\delta(t^{\prime})}e^{i\eta(t^{\prime})},
\end{eqnarray}
as well as the entanglement terms:
\begin{eqnarray}
    \delta \theta_{\cos} &=& \frac{1}{2}\int^{t_{g}}_{0}dt^{\prime}\cos(\omega t^{\prime})\frac{\Omega^{2}(t^{\prime})}{\delta^{2}(t^{\prime})} \nonumber \\
    \delta \theta_{\sin} &=& \frac{1}{2}\int^{t_{g}}_{0}dt^{\prime}\sin(\omega t^{\prime})\frac{\Omega^{2}(t^{\prime})}{\delta^{2}(t^{\prime})},
\end{eqnarray}
amounting to a Fourier cos and sin transform of $\gamma(t)$ and $|\gamma(t)|^{2}$. The results are shown in Fig.~\ref{fig:smooth_ff} in the main text. We can now see one of the biggest advantages of adiabatic gates, that, for important values of $\omega$ in the power spectral density, the time-averaged contribution is made increasingly small. Similar to Walsh sequences, the $\propto \exp(i\eta[t])$ component of $\alpha_{t,\cos(\sin)}$ averages to $0$ for low frequency noise, i.e. small $\omega$. In fact, unless $\omega\sim \delta(t)$ for a non-trivial length of time, the oscillatory behavior of the phase term will average $\alpha_{t,\eta}$ to zero, as it is a fast oscillating function that smoothly goes to zero at $t=0$ and $t=t_{g}$; this is not the case for Walsh functions, so, while they average $\omega\rightarrow 0$ noise to zero, we should not expect them to suppress the effect of time-dependent noise as well as adiabatic gates do. For the $\delta \theta_{\cos(\sin)}$ integrals, by contrast, $|\gamma(t)|^{2}$ does not time-average to zero\textemdash indicating adiabatic gates will be roughly as sensitive to entanglement errors as diabatic gates. 

\end{document}